\documentclass[journal]{IEEEtran}
\usepackage{amsmath}
\usepackage{cite}
\usepackage{graphicx}
\usepackage{array}
\usepackage{times}
\usepackage{subeqnar}

\begin{document}
\title{Mode-Locked Chip Laser using Waveguide Arrays}

\author{Xiao Zhang, Matthew Williams, Steven T. Cundiff and J. Nathan Kutz \thanks{X. Zhang is with the 
College of Applied Sciences, Beijing University of Technology, Beijing 100124, China.
M. Williams and J. N. Kutz are with the Department of Applied Mathematics, University of
     Washington, Box 352420, Seattle, WA 98195-2420.  Steven T. Cundiff is with JILA, University of Colorado and National Institute of Standards and Technology, Boulder, Colorado, 80309-0440 USA.  J. N. Kutz acknowledges
     support from the National Science Foundation (NSF) (DMS-1007621) and the U.S. Air Force Office of Scientific Research
(AFOSR) (FA9550-09-0174).}}%
 
\maketitle

\begin{abstract}
We demonstrate theoretically that robust mode-locking can be achieved on a semiconductor 
chip with a waveguide array architecture.  The waveguide arrays are used as an ideal saturable absorption
mechanism for initial noise start-up as well as pulse shaping and stabilization.  The cavity gain is
provided by an injection current and forward biasing of the semiconductor material.   The technology
can be integrated directly with semiconductor architectures and technologies, thus allowing for the potential of an 
on-chip, broadband device.
\end{abstract}


Mode-locked lasers are an increasingly important technological innovation as their potential applications have grown significantly over the past decade. Indeed, this promising photonic
 technology has a wide number of applications ranging from military devices and precision medical surgery to optical interconnection networks, broadband sources, and optical clocks. In some cases, such technologies have placed a premium on the engineering and optimization of mode-locked laser cavities that are aimed at producing output pulses of tens to hundreds of femtoseconds with maximal peak powers and energies.  Thus, the technological demand for novel techniques for producing and stabilizing 
ultra-short pulses has pushed mode-locked lasers, in particular fiber-based laser designs, to the forefront of commercially viable, nonlinear photonic devices.  Although fiber offers an attractive technological approach to mode-locking, it is envisioned that
chip-based devices could play a significant role if robust mode-locking can be demonstrated on-chip.  Here, we advocate the use of waveguide arrays for a chip-based mode-locked laser design and demonstrate that robust mode-locking can 
be achieved using the effective saturable absorption generated by the waveguides.

\begin{figure}[t]
\vspace*{-0.4in}
\centerline{\includegraphics[width=10cm]{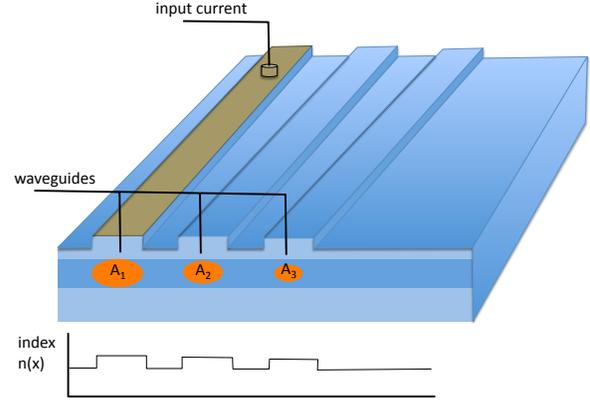}}
\vspace*{-0.3in}
\caption{Experimental configuration of chip-based waveguide array mode-locked laser.  Both saturable absorption~\cite{kutz_ol,kutz_jqe} and
amplification, via current injection and forward biasing of current of the semiconductor material~\cite{model1}, are accomplished in the waveguide array.  Only waveguide $A_1$ is amplified as waveguides $A_2$ and $A_3$ are used simply for saturable absorption and pulse shaping.  An A/R coating is highly desirable on both ends of the cavity
as the end mirror reflectivity is approximately 50\% without it.  The mode-locked field can be coupled out directly to
fiber~\cite{hudson}.}
\label{fig:setup}
\end{figure}

Optical waveguide arrays (WGAs) have been demonstrated to have a wide
range of potential photonic applications~\cite{review} including as the basis for all-optical signal
processing (routing and switching) in fiber optic networks and devices~\cite{christ,array1,array2,array3,array4}
and as potential saturable absorption mechanisms in mode-locked lasers~\cite{kutz_ol,kutz_jqe,kutz_bjorn,hudson}.
Indeed, with the advent of optical waveguide arrays, a new method
exists to generate robust, mode-locked pulses by using the ideal saturable absorption
behavior of the WGA~\cite{WAFFL}. 
Here the aim is to demonstrate that mode-locking can be achieved on a chip-based design, thus circumventing
the need for fiber and/or other external cavity components.  Figure~\ref{fig:setup}
demonstrates the envisioned experimental setup of a chip based mode-locked laser cavity.
Unlike previous studies of fiber-based passive WGA mode-locking~~\cite{kutz_ol,kutz_jqe,WAFFL}, the idea here is
to use current injection and forwarding biasing of the semiconductor material to produce the requisite gain for mode-locking.    Thus the all-chip cavity configuration envisioned takes advantage of several key physical effects for generating stable mode-locking.

%
%
%
%

 %
%
\begin{figure*}[t]
\centerline{\includegraphics[width=16cm]{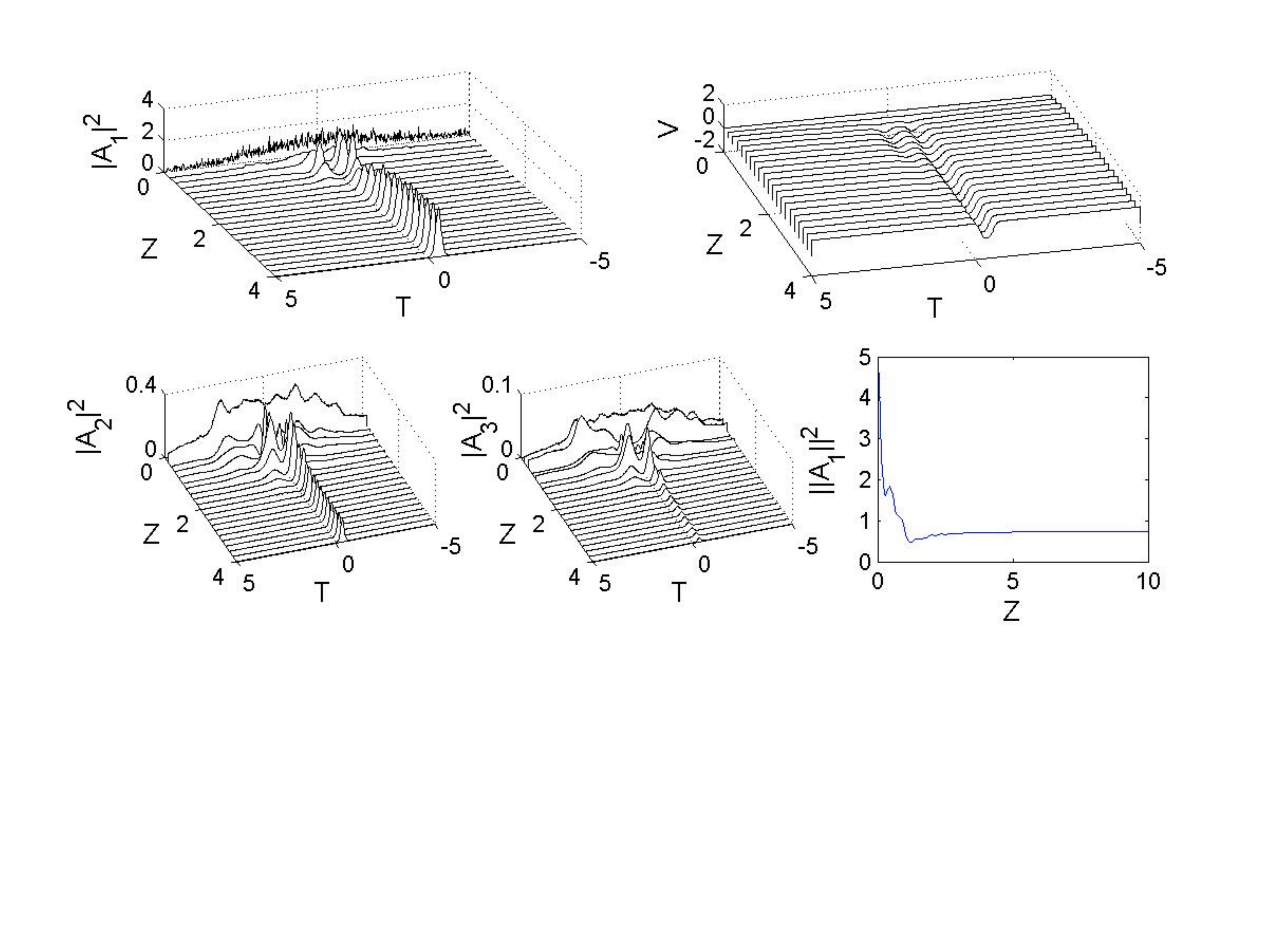}}
\vspace*{-1.7in}
\caption{Example of robust mode-locked chip dynamics. Here the three waveguide configuration of 
Fig.~\ref{fig:shape} is considered.  Only the first waveguide, $A_1$, experiences amplification from
the current injection and forward biasing of the semi-conductor material.  The top two panels demonstrate
the cold cavity start up of the electric field in the first waveguide along with the carrier density dynamics.  As
with typical mode-locked laser cavities, the mode-locked state acts as an attracting state of the system.  The
bottom panels demonstrate the electric field dynamics in the neighboring waveguides $A_2$ and $A_3$ along
with the energy equilibration in the first waveguide ($\| A_1 \|^2 = \int |A_1|^2 dT$).}
\label{fig:shape}
\end{figure*}
%
%

Semiconductor WGA lasers have been well studied by Rahman and Winful~\cite{model2} and others~\cite{model1} in
the context of continuous wave operation. Here, our aim will be to consider pulsed (mode-locked) operation of
the cavity.  Thus additional modeling terms will be required in the governing equations previously proposed~\cite{model1,model2}.
The governing equations for the propagation of an electromagnetic field in a semiconductor material 
arises from a combination of a modal field expansion and a high frequency asymptotic reduction of Maxwell's 
equations~\cite{model1,model2}. The evolution of the normalized electric field amplitudes in the waveguides $A_j$ and normalized carrier density in the first waveguide $V$ are given by
\begin{subeqnarray}
&& \hspace*{-.2in} \frac{\partial A_1}{\partial z}
=i(k_{11} A_1 + k_{12} A_2)
-i\frac{\beta^{''}}{2}\frac{\partial^2 A_1}{\partial t^2}
+i{\gamma}| A_1 | ^2 A_1
 \\
&& \vspace*{.5in}  \hspace*{.3in} -{\sigma} | A_1 | ^4 A_1
+(1-iR)V (1+{\tau}g \frac{\partial^2}{\partial t^2})A_1
-{\alpha}_1 A_1  \nonumber \\
&&  \hspace*{-.2in}\frac{\partial A_2}{\partial z}
=i(k_{12} a_1 + k_{22} A_2 + k_{23} A_3)
-{\alpha}_2 A_2\\
&& \hspace*{-.2in}\frac{\partial A_3}{\partial z}
=i(k_{23} A_2 + k_{33} A_3)
-\bar{\alpha}_3 A_3\\
&& \hspace*{-.2in}\frac{dV}{dz}
=\frac{1}{T} (P_1-(\xi_1+2 \gamma_1 V) |A_1|^2
-(1+\delta_1)V)
\label{eq:ge}
\end{subeqnarray}
where $z$ is the normalized distance in the waveguide array and $t$ is the normalized time.
The parameter $g$ represents the semiconductor gain strength which is determined by the current injection~\cite{model1}.
Note that gain is only applied to the first waveguide $A_1$ as shown in Fig. ~\ref{fig:setup}.
In the first equation, the parameters $\beta^{''}$, ${\gamma}$, ${\sigma}$, and ${\tau}$ determine the chromatic dispersion, nonlinear self-phase modulation generated from the Kerr nonlineartiy, the three-photon absorption 
coefficient~\cite{hud}, and the gain bandwidth in the semiconductor.  As is standard with coupled mode-theory, 
the parameters $k_{ij}$ measures the overlap of the modal structures from between waveguide $i$ and $j$~\cite{yariv}.  
For the carrier density, 
the parameters $R$, $T$, $\xi$, $\gamma$, $\delta$ are related to the time-response, size and specific semiconductor
material used~\cite{model1,model2}.  

For AlGaAs, for instance, many of the parameters can be explicitly 
characterized.  For instance, it is known that the chromatic dispersion coefficient is 1.25~ps$^2$/m, the
nonlinear index is 3.6~m$^{-1}$W$^{-1}$
and that the three-photon absorption coefficient is approximately 2$\times$10$^{-5}$~m$^{-1}$W$^{-2}$~\cite{hud}.
However, many of the other coefficients depend largely on the waveguide separation and waveguide width, parameters
that can be tailored and engineered for the mode-locking requirements here.  Waveguide widths can be easily
envisioned to be on the order to 1-5~microns with separations on the order of 2-10~microns.  A driving current
for the system would be somewhere in the tens of milliamps regime~\cite{model1}.   Given the large parameter space
to be potentially explored, our computational studies will focus on simply demonstrating the concept and potential ability to construct such a mode-locked chip device.  Clearly the choice of specific semiconductor material, its time constants, nonlinearity, and geometrical configuration give a great deal of flexibility in achieving the parameter regime necessary
for mode-locking.

To demonstrate the mode-locking process, Eqs.~(\ref{eq:ge}) are simulated over a large number of round trips.
From a cold cavity startup, i.e. an initial low-amplitude, white-noise electromagnetic field, the cavity quickly forms
into a robust mode-locked pulse.  Figure~\ref{fig:shape} demonstrates the mode-locking process for a white-noise
initial condition.  As is expected in a mode-locked cavity, the effective saturable absorption of the WGA quickly shapes
the pulse into a robust mode-locked pulses while also stabilizing the cavity energy.  The carrier density also takes on
a robust temporal response during the mode-locking process.  As with a fiber based device~\cite{kutz_jqe}, the 
electric field in the neighboring waveguides $A_2$ and $A_3$ inherit their from $A_1$.  Indeed, both waveguides
$A_2$ and $A_3$ experience a net loss.  However, $A_1$ continuously feeds energy into these waveguides, thus 
stabilizing the robust structures shown.  The bottom right panel shows the equilibration of cavity energy during the
mode-locking process.
For this demonstration, the dimensionless parameters in Eqs.~(\ref{eq:ge}) are given by $ k_{11}=k_{22}=k_{33}=0, k_{12}=k_{23}=6,\beta^{''}=-1, \bar{\gamma}=20,\bar{\sigma}=0.1,\bar{\tau}=0.1,R=3,T=1,\xi_1=1,\gamma_1=1,
\delta_1=0, \bar{\alpha}_1=\bar{\alpha}_2=0,\bar{\alpha}_3=10.$
If $v_1>0,$ then $g=1.$ If $v_1<0,$ then $g=0.$

%
%
\begin{figure*}[t]
\centerline{\includegraphics[width=15.0cm]{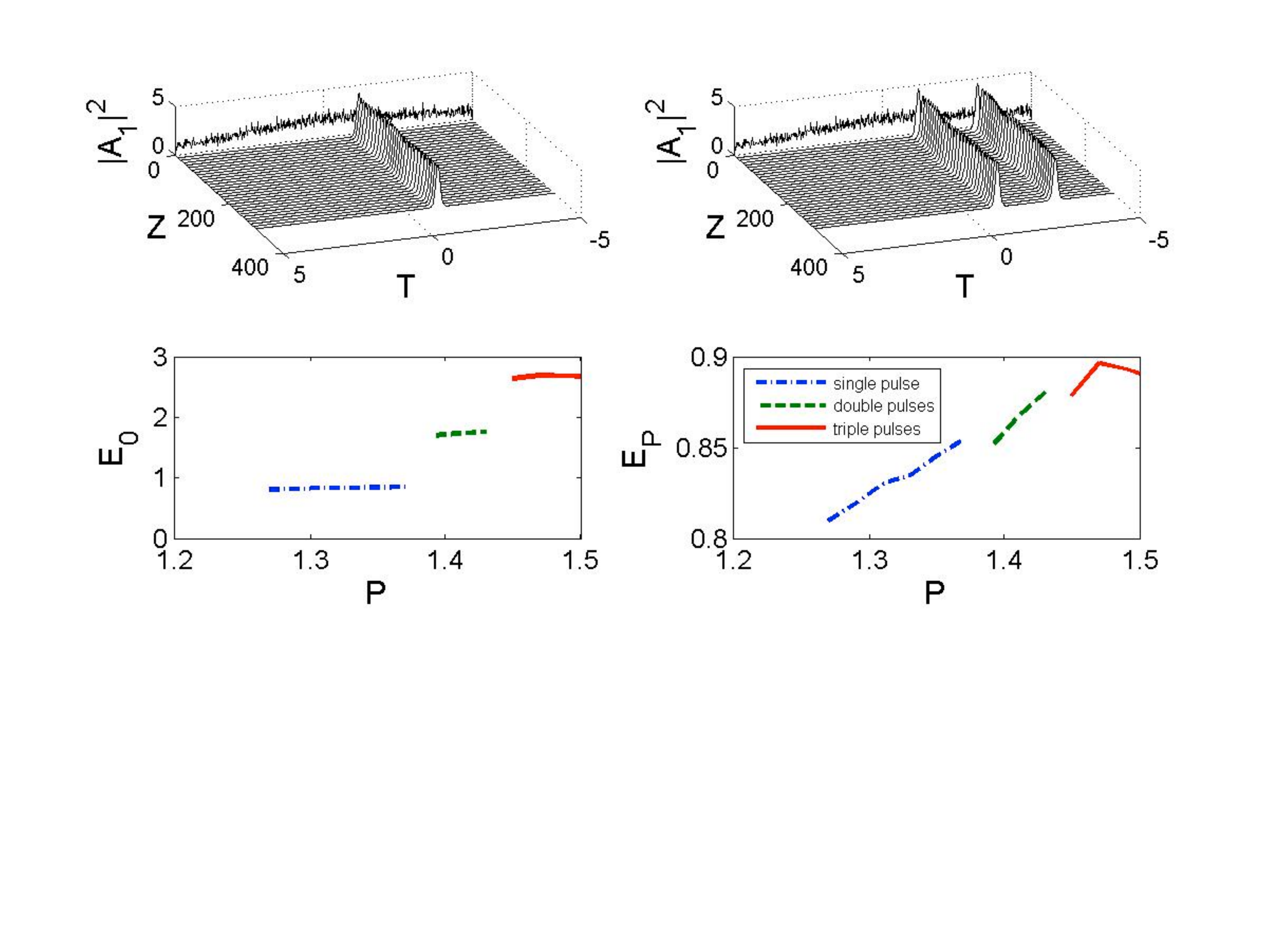}}
\vspace*{-1.6in}
\caption{Mode-locked chip dynamics as a function of increased injection current (gain) on waveguide
 $A_1$.  The top left panel shows the single pulse per round trip configuration that is desirable whereas
 the top right panel shows the result of increasing the injection current, thus producing higher gain and two pulses per 
 round trip.  The bottom panels shows the cavity energy for the total cavity, $E_0$, (left bottom) and for
 each individual pulse, $E_p$ (right bottom).}
\label{fig:modelock}
\end{figure*}
%
%

As with all mode-locked lasers, the performance limits of single-pulse operation are typically always limited by
gain bandwidth restrictions.  Thus as the injection current is increased, single pulse operation as illustrated
in Fig.~\ref{fig:setup} is destabilized and a multi-pulsing transition is exhibited.  Such a phenomenon is
generic for laser systems that are driven by the interplay of saturable gain and nonlinear losses~\cite{multi}.
This is also the case for the WGA chip laser.  In particular, as the gain is increased through ramping up
of the injection current, the single-pulse per round trip bifurcates the a two-pulses per round trip configuration.
Figure~\ref{fig:modelock} demonstrates this bifurcation process.  The top two panels demonstrate the robust 
single- and double-pulse per round trip configurations that arise from an initial noise startup.  The bottom panels
show the pulse energy for the entire cavity, i.e. the sum of all the pulses, along with the energy of
each individual pulse during the bifurcation process from one to three pulses per round trip.  On the one hand,
the multi-pulsing instability is often deleterious as it limits the energy and power of single pulse operation.  On the
other hand, the behavior suggests that the chip laser is in keeping with the fundamental behavior of other mode-locking
configurations, thus potentially allowing one to apply the intuition of cavity design to the chip laser.

In conclusion, we have demonstrated theoretically a proof-of-concept mode-locked chip laser.  Thus robust, ultra-short pulse
generation can be achieved on an all-chip device, potentially revolutionizing the use of
mode-locked lasers in semiconductor architectures.  The mode-locked chip laser is based upon
proven WGA technology and its previously demonstrated saturable absorption properties in
a fiber-based setting~\cite{WAFFL}.  Given its technological promise, the chip laser is a promising
advancement for integration into chip design and architecture.  It only remains to pick appropriate semiconductor
material and its geometrical configuration in order to access a parameter space where robust mode-locking 
occurs.  The primary hurdle in achieving this physical regime is the typical time constants associated with the recover
of the carrier dynamics.  However, longer cavities can be generated on chip (See Fig.~\ref{fig:setup}). In practice, managing the cavity losses has been shown in practice to be very important
in achieving mode-locking~\cite{WAFFL}.     Moreover
cavity design (waveguide width and spacing) along with an appropriate semiconductor material may be able
to render this a robust and inexpensive mode-locking technology that can greatly impact semiconductor laser technologies.


\bibliographystyle{IEEEtrans}




















\end{document}